# Specific, cumulative and reversible odour learning in a whole-brain connectome model of *Drosophila*

Daniel Asis, Independent Researcher, Parañaque City, Philippines. ORCID: https://orcid.org/0009-0009-1456-5276. Correspondence: danielasis.research@gmail.com.

## Abstract

Whole-brain models of the fruit fly built from the FlyWire connectome can predict how neurons respond to a stimulus, but they keep every synapse fixed and so cannot learn, while models that do learn simulate only the mushroom body and feed it with artificial odour signals. In this paper, I made one type of synapse plastic in a whole-brain leaky integrate-and-fire model of the FlyWire connectome: the synapses from Kenyon cells (KCs), those that encode odours, to mushroom body output neurons (MBONs), those that carry the fly's response to them. Under the learning rule, dopamine weakens each of these synapses in proportion to how strongly its KC fired, matching the weakening measured in living flies. Odours reached the KCs through the connectome's mapped wiring from the olfactory receptor neurons, but because the model has no sensors for pain, punishment was instead delivered by driving the 12 PPL1 dopamine neurons with random spikes at 100 Hz during the odour. Pairing an odour with punishment lowered the response of approach MBONs to that odour from 11.3 to 0.3 Hz in all five noise replicates of one simulated brain, while an unpaired odour, a never-paired odour and control runs with the rule switched off or scrambled changed little. When a second odour was paired on the brain already trained on a previous odour, it was learned at 85.5% of the strength reached by a naive brain, and at the end of training the approach MBONs still responded weakly to both odours, at 1.6 and 1.1 Hz against about 11 and 12 Hz before training. With a second rule, in which dopamine arriving before the odour strengthens a synapse instead of weakening it, a simulated fly body driven by the model unlearned and relearned in four of four cycles, whereas the first rule got stuck after one; the same potentiation, tested with backward pairing alone rather than as part of a full cycle, was too small to reach behaviour on its own. All results are for aversive, punishment-driven learning; reward learning could not be tested because, at the synaptic weight and synapse floor used here, sugar input does not recruit the reward dopamine neurons in this model. Learning depended on an authored plasticity rule, an injected punishment signal, and several other authored model choices, detailed in Methods; given these choices, the odour-specificity of learning followed from the structure of the connectome.

Keywords: *Drosophila*; connectome; mushroom body; synaptic plasticity; olfactory conditioning; whole-brain simulation

## Introduction

The first complete wiring diagram of an adult fly brain, the FlyWire connectome, maps about 139,000 neurons and the synapses between them (Dorkenwald et al., 2024; Schlegel et al., 2024). Shiu et al. (2024) turned this map into a whole-brain model in which every neuron is a leaky integrate-and-fire (LIF) unit, and many of the model's predictions about feeding and grooming circuits were then confirmed in living flies. The same model has since been used to trace sweet taste to the brain's reward neurons (Christie et al., 2026). In all of this work, every synaptic weight stays fixed, so the model cannot change with experience, and a brain model that cannot learn cannot be used to test theories of memory.

In living flies, odour memories form in the mushroom body. Each odour activates a sparse set of Kenyon cells (KCs; Honegger et al., 2011; Turner et al., 2008), and when dopamine arrives together with the odour, it weakens the synapses from those KCs onto mushroom body output neurons (MBONs), so the output to that odour changes (Hige et al., 2015). Computational models have captured this learning, but they simulate only the mushroom body and feed it with artificial odour signals, such as random sets of active KCs (Bennett et al., 2021; Jiang & Litwin-Kumar, 2021). These models therefore cannot show whether the fly's real wiring, from the olfactory receptor neurons through the antennal lobe to the KCs, delivers odour signals that the mushroom body can learn from, which is the step that Jiang and Litwin-Kumar (2021) named for future work.

Several public codebases have begun to add learning to models built from these connectomes. One simulates the entire nervous system of a male fly with plasticity at the synapses from KCs to MBONs and reports that the learned change reverses when the punished odour is swapped (TheMrRaGe, 2026). Another weakens the synapses from KCs to MBONs in the whole-brain LIF model of Shiu et al. (2024) after an odour is paired with punishment, and reports a larger drop in MBON response to the paired odour than to a control odour, although in its main experiment the odour is delivered by driving a chosen set of KCs directly (Simpbuilder, 2026). A third adds a rule that depends on the timing of odour and dopamine, of the kind measured by Handler et al. (2019), to a network of about 5,600 mushroom body neurons taken from FlyWire, again with odours delivered directly to the KCs (Turayev, 2026). A fourth adds dopamine-gated changes to the FlyWire model, applied in part to the output rates after each simulation rather than to the synapses within it, and reports no quantitative learning results (lixiang1076, 2026). None of these has been written up as a study, and none asks whether a trained whole-brain model can learn a second odour without erasing the first, or unlearn and relearn the same odour.

Here I ask how specific, cumulative, and reversible the learning of an odour paired with punishment is, in a whole-brain LIF model of the FlyWire connectome, when plasticity is limited to the synapses from KCs to MBONs. The measured learning in this model depends on an authored plasticity rule, an injected punishment signal, and the other authored components listed in Table 1. I set every global constant once without training it on the learning tasks. Given these components, its specificity to the paired odour follows from the structure of the connectome. I find that learning stays specific to the paired odour, that a second learned memory forms on an already-trained brain at 85.5% of the strength reached by a naive one, and that a simulated fly body driven by the model unlearns and relearns in four of four cycles when the rule depends on the timing of odour and dopamine.

## Methods

**Connectome and neuron model.** I simulated the whole brain of an adult female fly using version 783 of the FlyWire connectome (Dorkenwald et al., 2024), with cell types from Schlegel et al. (2024); the resulting network has 139,248 neurons. Each neuron was a leaky integrate-and-fire unit with the parameters of Shiu et al. (2024): a resting and reset potential of −52 mV, a spike threshold of −45 mV, a membrane time constant of 20 ms, a synaptic time constant of 5 ms, a refractory period of 2.2 ms, a synaptic delay of 1.8 ms, and a weight of 0.275 mV per synapse, with the sign of each connection set by the predicted neurotransmitter of the sending neuron (Eckstein et al., 2024). I ported the model to a GPU with a time step of 0.1 ms, with the port changing memory layout and batching, but not the equations. Unlike the released model of Shiu et al. (2024), which keeps every connection, I set a minimum floor of 5 synapses per connection, which meant every connection made of fewer than 5 synapses was dropped. Supplementary Table S1 lists which results change when this floor is lowered to 1. At this floor, with the transmitter signs used throughout this paper (Shiu et al.'s predicted signs, except the 44 antennal-lobe local neurons described in Methods, whose outgoing sign is flipped), the model reproduced 8 of 13 benchmark results from Shiu et al. (2024) fully, 3 partly, and failed 2. One failure probably comes from the floor, which weakens the response to weak sugar input; the other remains unexplained. As a check on the floor specifically, I also ran the model as Shiu et al. (2024) released it — their connection matrix at a floor of 1, with their original transmitter signs — on 5 of the 13 benchmarks; because this configuration differs from the canonical one in both floor and sign at once, it does not separate the two effects. It passed 4 of the 5 and failed 1: a 5.2 Hz drop in one motor-neuron readout between 180 and 190 Hz, within the roughly 4-6 Hz per-seed spread at those rates but outside the 2 Hz preregistered tolerance. The other 8 benchmarks were not run at this configuration.

**Authored components.** Table 1 lists every authored component that comes from neither the connectome nor Shiu et al. (2024). I chose each one from checks run before the learning tests reported here, and did not change it after seeing their results. Two of them change how the brain processes odour. The antennal lobe contains 44 excitatory local neurons that release acetylcholine (Shang et al., 2007); in the fly, excitation between glomeruli also travels through electrical coupling (Yaksi & Wilson, 2010), which a model with only chemical synapses cannot express. Left as chemical synapses, these neurons merged every odour into one response: when I drove eight odour channels one at a time at 80 Hz, the sets of projection neurons that each pair of odours activated overlapped with a mean Jaccard similarity of 0.961. I negated their outgoing weights, which lowered the overlap to 0.057 but left only 0.7% of KCs active, too few to tell odours apart, so I multiplied the weights from projection neurons to KCs by 8. With a gain of 6, both tests still showed learning specific to the paired odour, but the paired-odour (DC2) shift was roughly half as large in each test (5.08 vs 10.96 Hz; 5.10 vs 10.90 Hz), the second odour in the accumulation test was learned much less completely (a shift of 2.58 vs 9.66 Hz, about a quarter), and the small shift for the unpaired odour in the specificity test collapsed to near zero (Table S2).

**Table 1.**

| Component | Setting | Reason | Reference |
|---|---|---|---|
| Synapse floor | Connections with < 5 synapses dropped | Choice; released model uses 1 (Table S1) | Shiu et al., 2024 |
| AL local neurons | Outgoing weights of 44 excitatory local neurons negated | Electrical coupling not expressible; merged all odours | Shang et al., 2007; Yaksi & Wilson, 2010 |
| PN → KC gain | × 8 | Restores KC recruitment; at gain 6, the paired-odour shift roughly halves and the second odour is learned much less (Table S2) | — |
| Odour | One receptor channel at 80 Hz for 300 ms | Stands in for an odour | — |
| Odour channels | DC2 paired, D unpaired, DA1 never paired | DA1 excluded from pairing: innate pheromone meaning | Kurtovic et al., 2007 |
| Punishment | PPL1 dopamine neurons driven at 100 Hz (60 Hz in the closed loop) | No pain or shock sensors in the model | Claridge-Chang et al., 2009 |
| Plastic synapses | 18,674 KC → MBON connections; each MBON labelled approach or avoid | Where learning happens in flies | Aso et al., 2014a, 2014b; Hige et al., 2015 |
| Learning constants | $\eta = 5 \times 10^{-6}$; $\lambda = 0.01$; weights kept in [0.1, 1.8] × baseline | Set once before any run | Bennett et al., 2021; Yamada et al., 2024 |
| Timing rule | Odour before or with dopamine weakens; dopamine before odour strengthens | Relearning test only | Handler et al., 2019 |
| Body links | Odour → receptor rate (logarithmic); descending neurons → leg rhythm; virtual corridor | Stand-ins for the nose, nerve cord and arena | Wang-Chen et al., 2024 |

**Odour and punishment.** I presented an odour by driving one class of olfactory receptor neurons (ORNs) with Poisson spikes at 80 Hz for a 300 ms trial: the ORNs of glomerulus DC2 (20 neurons) for the paired odour, those of glomerulus D (31) for the unpaired odour, and those of glomerulus DA1 (126) as a probe that was never paired. I kept DA1 out of pairing because its ORNs detect the pheromone cVA, to which flies respond without learning (Kurtovic et al., 2007), so a change in its response could not be told apart from an innate one. Before any learning run, a gate script confirmed that each channel activated 5.5–8.9% of KCs, that DC2 and D shared few active KCs (Jaccard similarity 0.057), and that no KC fired without odour. I delivered punishment by driving the 12 PPL1 dopamine neurons with Poisson spikes at 100 Hz during the same 300 ms as the paired odour, since activating these neurons works as punishment in flies (Claridge-Chang et al., 2009) and the model has no sensors for pain or shock.

**Learning rules.** Only the 18,674 synapses from KCs to MBONs could change. I labelled each MBON as approach or avoid following Aso et al. (2014b), and punishment acted only on synapses onto approach MBONs, so any change to avoid MBONs had to travel through the rest of the network. The dopamine signal was the mean firing rate of the 12 PPL1 dopamine neurons, applied to every approach synapse rather than routed by compartment. For the specificity and accumulation tests, I used a rule that only weakens synapses (Hige et al., 2015), applied once per paired trial: each weight fell by $\eta \cdot e \cdot d \cdot w_0$, where e is the firing rate of the KC, d is the dopamine rate, $w_0$ is the weight from the connectome, and $\eta = 5 \times 10^{-6}$ per $Hz^2$. Weights could not fall below 0.1 $w_0$, and after each update they recovered toward $w_0$ by a fraction $\lambda = 0.01$, the constant strengthening that Bennett et al. (2021) added to keep this kind of rule from saturating. For the relearning test, I used a rule based on timing (Handler et al., 2019), applied every 100 ms: dopamine that arrived with or after KC activity weakened the synapse, dopamine that arrived before it strengthened the synapse, and weights stayed between 0.1 and 1.8 $w_0$, an upper bound of the same order as the long-term strengthening measured by Yamada et al. (2024). I set η in an earlier pilot as the value at which the weights of every seed were weakened partly, rather than all the way to the floor or not at all, and I used the same η, λ and bounds in every test. Unlike the synapse floor and the PN-to-KC gain, η was not swept for robustness, so how sensitive the results are to this choice is untested.

**Specificity test.** A probe presented each of the three odours 5 times with no punishment and no learning, and I probed the brain once before training and again after every 5 training trials. A training trial presented the paired odour with punishment and then the unpaired odour without it, each for 300 ms in a separate simulation that started from rest, and the weights changed after the paired presentation only. Each run had 30 training trials. I ran four arms with 5 seeds each: learning, lesion, shuffle and reversed. The learning arm followed this protocol. The lesion arm repeated the same trials with the learning rule switched off, to test whether the change needs plasticity. The shuffle arm kept the rule on but made it read each synapse's activity from a KC chosen at random, to test whether the change needs the synapse's own KC to be active. The reversed arm paired D instead of DC2, to test whether the result depends on which odour is paired.

**Accumulation test.** I trained one brain in three phases without resetting its weights: 30 trials pairing DC2 with punishment, a pause of 30 trials with no odour and no punishment, during which only the recovery term acted, and 30 trials pairing D with punishment. I ran three arms with 5 seeds each: interfere, retain and lesion. The interfere arm ran all three phases. The retain arm ran only the first two phases, to measure how much of the DC2 change the pause alone removes. The lesion arm ran all three phases with the rule switched off, as in the specificity test. As the reference for how well D is learned from scratch, I used the reversed arm of the specificity test, which trained D on a naive brain.

**Relearning test.** The brain drove the NeuroMechFly v2 body (Wang-Chen et al., 2024), which walked on a simulated ball in a virtual corridor. A single source of an odour mixture drove the olfactory receptor neurons of both DC2 and D, more strongly as the fly came closer; when the fly came within 1.0 mm, the source jumped back to 2.5 mm and I counted an arrival. The brain and body exchanged signals every 100 ms, a step I call a tick. In a paired block, the 12 PPL1 dopamine neurons were driven at 60 Hz on the tick after each arrival, so dopamine followed the odour. In a backward block, the odour was withheld on the tick of dopamine and the tick before it and returned after it, so dopamine came before the odour, and the number of punishments matched that of the paired block. A test block lasted 60 ticks, with the weights frozen, no punishment, and the brain, body and position reset. Each chain began with a test and then ran four cycles of a 200-tick paired block, a test, a 200-tick backward block and another test. I ran this chain for both rules with the same 5 seeds. In a separate set of shorter arms, with 80 training and 60 test ticks, I compared paired, backward and unpaired punishment against a lesion arm; in the unpaired arm, punishment was delivered at the far end of the corridor.

**Readouts and statistics.** In the specificity and accumulation tests, the readout was the avoidance index: the difference between the mean firing rates of avoid and approach MBONs during a probe, so the index rises as the odour loses its pull on approach MBONs. Learning was measured by the change in avoidance index from the baseline to the latest probe. I also report the two groups of MBONs separately, because the index can rise through either one. In the relearning test, the readout was walking speed: the forward distance the fly covered in the virtual corridor divided by the length of the test block. Seeds are noise replicates of one simulated brain, not separate brain instances, so I report the mean with the standard deviation (SD) across seeds and the number of seeds whose change had the same sign, and I do not report p-values. For the separate relearning arms, the preregistered bar was a difference from the lesion arm of at least 3 standard errors of the mean (SEM), with the same sign in 5 of 5 seeds for paired punishment and in at least 4 of 5 for backward pairing. Only the rules for the relearning test were written into a dated, version-controlled file before the test ran; the rules for the specificity and accumulation tests are in planning notes and code that were committed together with their results, so their timing cannot be checked from the record.

**Code and data availability.** The simulation code is available under the MIT licence at https://github.com/AllMites/neuroplasticfly (release v0.1). A single command reproduces the specificity test at seed 0; on a fresh download, it reproduced the paired shift at +10.90 Hz against the 5-seed mean of +11.05 Hz reported here. The connectome is not redistributed; the code downloads it from its public sources, the connectivity data of the FlyWire Consortium (2024) on Zenodo and the cell annotations in the flywire_annotations repository (flyconnectome, 2026), which draws on the visual-system wiring diagram of Matsliah et al. (2024) and the male-nervous-system connectome of Berg et al. (2026), under the terms set by FlyWire. At release v0.1, the specificity and accumulation tests are reproducible from the public code; the closed-loop relearning pipeline, which drives the NeuroMechFly v2 body, is not included and so its result cannot yet be independently checked. The repeat of the protocol of Christie et al. (2026) described in the Discussion, with its preregistration, is also in the repository.

## Results

**Learning was specific to the paired odour.** After 30 training trials, the avoidance index for the paired odour, DC2, had risen by 11.05 ± 0.20 Hz (mean ± SD), and it rose in all five seeds, noise replicates of one simulated brain, not independent brains (Fig. 2a). Over the same trials, the index for the unpaired odour, D, rose by only 1.43 Hz, and the index for the never-paired odour, DA1, by 0.56 Hz. Even the small change for D depended on the gain from projection neurons to KCs: in seed 0, it fell from 13.9% of the paired change to almost nothing when I lowered the gain from 8 to 6 (Table S2). This specificity was tested with one pair of channels, DC2 against D, plus DA1 as a never-paired probe; whether it holds for a larger panel of odours, including ones whose receptor-neuron representations overlap more than DC2 and D happen to, is untested.

The control arms showed that the change needed both the rule and the correct pairing. With the learning rule switched off, the index for the paired odour moved by −0.73 Hz, and with the rule reading randomly reassigned KCs, it moved by 0.54 Hz, which is 4.9% of the change in the learning arm. When D was paired instead of DC2, the result was the mirror image: D rose by 11.29 Hz and DC2 by only 1.74 Hz.

At this synapse floor, the whole change came from the approach MBONs, whose response to the paired odour fell from 11.3 to 0.3 Hz while the avoid MBONs barely moved (Fig. 2b). At a floor of 1, the avoid MBONs also rose, by 4.2 Hz against 0.1 Hz at a floor of 5, and the shifts for the paired odours were 2.1 to 2.7 times larger, for example +10.96 Hz to +29.25 Hz for the DC2 shift, measured the same way as in the accumulation test below; this is not a small correction, since it changes which class of MBON the effect is attributed to (Table S1). In the learning arm, the index for the paired odour rose steadily over training and levelled off by about the twentieth trial (Fig. 2c).

**A second memory formed on an already-trained brain without erasing the first.** The first phase reproduced the specificity test, raising the avoidance index for DC2 by 10.96 ± 0.27 Hz. When I then paired D with punishment on this trained brain, its index rose by 9.66 ± 0.45 Hz, which is 85.5% of the 11.29 Hz that a naive brain reached, while in the lesion arm it hardly moved (Fig. 3a). By the end, both odours had lost their pull on the approach MBONs, which now responded at 1.6 Hz to DC2 and 1.1 Hz to D instead of about 11 and 12 Hz. The weakening of the plastic synapses added up rather than being replaced: its mean nearly doubled, from 1.5% after the pause to 2.8% after the second training (Fig. 3b).

Learning the second odour did cost the first one something. Across 5 seeds, DC2 lost 1.12 Hz of its change while D was trained, 5.7 times the 0.20 Hz it lost over an equal stretch without training, but it still kept 88% of its change. In seed 0, this interference depended on the synapse floor, and not just in size: at a floor of 5, training D cost DC2 2.3 times what an equal stretch without training did (against the 5-seed ratio of 5.7 above); at a floor of 1, it cost only 0.67 times as much, and DC2 kept 69% of its change, against 89% at floor 5 (both seed 0). Whether the second memory interferes with the first, not only how much, is therefore a property of this modelling choice rather than a settled property of the fly (Table S1).

The stretch without training is not evidence of forgetting. Over those 30 trials, the synapses kept 74.0% of their weakening, exactly the value of $0.99^{30}$ that the recovery constant λ sets, so this number checks the code rather than measuring a property of the brain. The avoidance index kept 98.2% over the same stretch, which overstates the retention, because the approach MBONs stayed so close to zero that the index could not show their recovery.

**Only the rule with timing relearned repeatedly.** When the simulated fly was punished on reaching the odour, it walked more slowly than the fly in the lesion arm, by 0.28 mm/s, a difference of 13.7 SEM that appeared in all five seeds. Punishment given away from the odour, at the far end of the corridor, slowed the fly by only 0.04 mm/s (2.2 SEM), even though this unpaired arm received four to five times as many punishments. Backward pairing on its own, in which dopamine came before the odour, did not make the fly walk faster than the lesion arm, so it failed its preregistered bar.

The cycles of learning and unlearning separated the two rules (Fig. 4a). Under both rules, the first block of paired punishment slowed the fly from 2.14 mm/s to about 0.2 mm/s. With the timing rule, every backward block then brought the fly back to between 2.07 and 2.10 mm/s, on average 97% of its naive speed, and every paired block slowed it down again; after four cycles, the swing was as large as it had been after the first. With the rule that only weakens synapses, the first backward block brought the fly back only to 0.95 mm/s, consistent with the passive recovery that the constant λ alone would be expected to produce over that block (as in the weight-retention check in the accumulation test above), though this was not separately confirmed here with a lesion-style control; in the three later cycles the fly never walked faster than 0.32 mm/s.

Two points qualify this result. First, the preregistered rule judged relearning by the direction of each change rather than by its size, so it labelled both rules as relearning; by size, only the timing rule relearned, and the weakening-only rule's partial recovery in the first backward block is consistent with passive recovery from λ rather than new relearning, on the same reasoning as above. Second, at the reference setting the drive to the legs was at its maximum at 75% of time points, so the fly's speed flipped between nearly full and nearly stopped, which could have hidden a graded effect behind a ceiling. To check this, I reran the closed loop with the descending-neuron reference rate that sets the gain from firing rate to leg central-pattern-generator (CPG) amplitude raised from 1.0 to 1.9, which lowers that gain and gives the readout more headroom before saturating, at the cost of a slower naive speed (1.355 mm/s instead of 2.14 mm/s): the timing rule again relearned in all four cycles (Fig. 4b), and backward pairing on its own still failed its bar, so the timing effect does not depend on the readout being saturated.

## Discussion

A whole-brain model of the fly, in which every synapse kept its connectome weight except those from KCs to MBONs, learned to stop treating a punished odour as attractive, and this learning was specific to the paired odour, cumulative across two odours and, with a rule based on timing, reversible. Because the odours reached the KCs through the mapped wiring rather than as artificial patterns, the specificity reflects what that wiring delivers: once two authored changes had restored separate odour codes in the antennal lobe and the KCs, a single plasticity rule was enough to tell a punished odour from an unpunished one. The model is therefore a test bed in which theories of odour memory can be run on the fly's real input pathway.

These results agree with the odour-specific weakening of synapses from KCs to MBONs that Hige et al. (2015) measured in living flies, and they extend models of the mushroom body that used artificial input (Bennett et al., 2021; Jiang & Litwin-Kumar, 2021). The rule that only weakens synapses got stuck after one cycle of relearning, which fits the limited learning that Bennett et al. (2021) reported for rules of this kind, and the timing rule of Handler et al. (2019) removed that limit. A rule of this kind has also been added to a mushroom body network taken from FlyWire, where dopamine arriving before the odour strengthened the synapses after a single pairing (Turayev, 2026); here the rule was tested over repeated cycles of learning and unlearning in a whole-brain model. An independent whole-brain model ran into the same failure in the antennal lobe, where local neurons merged every odour into one response, and it also fixed the failure by changing the sign of local neurons, although it re-signed a larger set of 151 cells on the grounds that their transmitter had been misclassified (TheMrRaGe, 2026). That two separate models met the same failure suggests it comes from simulating the antennal lobe with chemical synapses only, not from either codebase.

The results have clear limits. The plasticity rule, the punishment signal and several settings are authored, and I injected the punishment at the PPL1 dopamine neurons because the connectome model has no sensors for pain or shock. All results come from five noise replicates of one simulated brain, not from separate brain instances, so they show that the phenomenon is reproducible under stochastic variation within this one connectome, not that it holds across independently reconstructed brains. Most consequentially, the choice of synapse floor changes which conclusions hold: at the 5-synapse floor, the change appeared only at the approach MBONs, partly by construction, because punishment acted only on their synapses; the second memory also cost the first one more than an equal stretch without training did, but at a floor of 1 the avoid MBONs changed too and that comparison flipped, so that training the second odour then cost the first memory less than an equal stretch without training. A reader should not take either the approach-only result or the direction of that cost comparison as a property of the fly brain itself, as both heavily depend on the implementation choice. The walking readout was close to all-or-nothing, so the relearning test shows that the fly unlearned and relearned, not by how much. The stretch without training measured the recovery constant I set, not forgetting. Finally, reward learning was not tested, because sugar input does not recruit the reward dopamine neurons in this model: with a weight of 0.275 mV per synapse and the 5-synapse floor, none of the 307 PAM dopamine neurons fired. Christie et al. (2026) used the same model with stronger synapses and report that, at a weight of 0.39 mV and high rates of sugar input, most of these neurons respond, through a pathway that runs from the sugar receptor neurons through a neuron they call Fox, and in their simulations, silencing Fox removes the response.

When I repeated their protocol, driving the sugar receptor neurons and gustatory interneurons of one side for 30 trials of 1000 ms at weights from 0.37 to 0.39 mV, at most 17 of the 307 PAM neurons responded with a floor of 1 and at most 14 with a floor of 5, compared with the roughly 200 they report. Silencing Fox did not lower the count, which suggests that the few responses I saw came from bursts of activity across the whole brain rather than from the Fox pathway.

The difference may lie in a setting I could not match, such as the synapse floor, which their paper does not specify, so this shows that the response did not appear in my implementation, not that it is absent from the model. Lowering the floor to 1 did not restore it at any of these weights, so the floor's weakening of the response to weak sugar input does not by itself explain the failure. With the settings used here, the same failure also appears when the neurons are modelled as passive compartments or as calibrated rate units.

The model makes predictions that can be tested in flies with calcium imaging. After an odour is paired with activation of PPL1 dopamine neurons, the response of approach MBONs should fall for that odour but not for an unpaired odour that activates few of the same KCs, and a second pairing with another odour should lower the response to that odour while the response to the first stays low. The model can also be extended with mechanisms it now lacks, such as electrical coupling in the antennal lobe, neuromodulators other than dopamine, and activity that persists after a stimulus ends, each added as an optional, disclosed module and tested against the results reported here.

In this FlyWire-derived model, with the authored components and parameters stated above, restricting plasticity to the synapses from KCs to MBONs was enough to produce odour-specific learning and a second learned odour on top of the first, and the places where the model falls short point to the mechanisms that a learning brain model needs next.

**Acknowledgements.** I thank Lisha Shao for helpful correspondence.

**Use of AI tools.** The simulation code and analysis scripts were written with the help of an AI coding assistant (Claude, Anthropic), and an AI writing assistant helped draft and edit the text. I designed the study, chose every parameter and control, set the pass/fail rules, checked every reported number against the result files, and I take full responsibility for the content.

**Competing interests.** The author runs a public social media channel that features simulations built on the same whole-brain model and may feature this work; the channel has no funding tied to these results.

**Funding.** This work received no funding.

# Figures

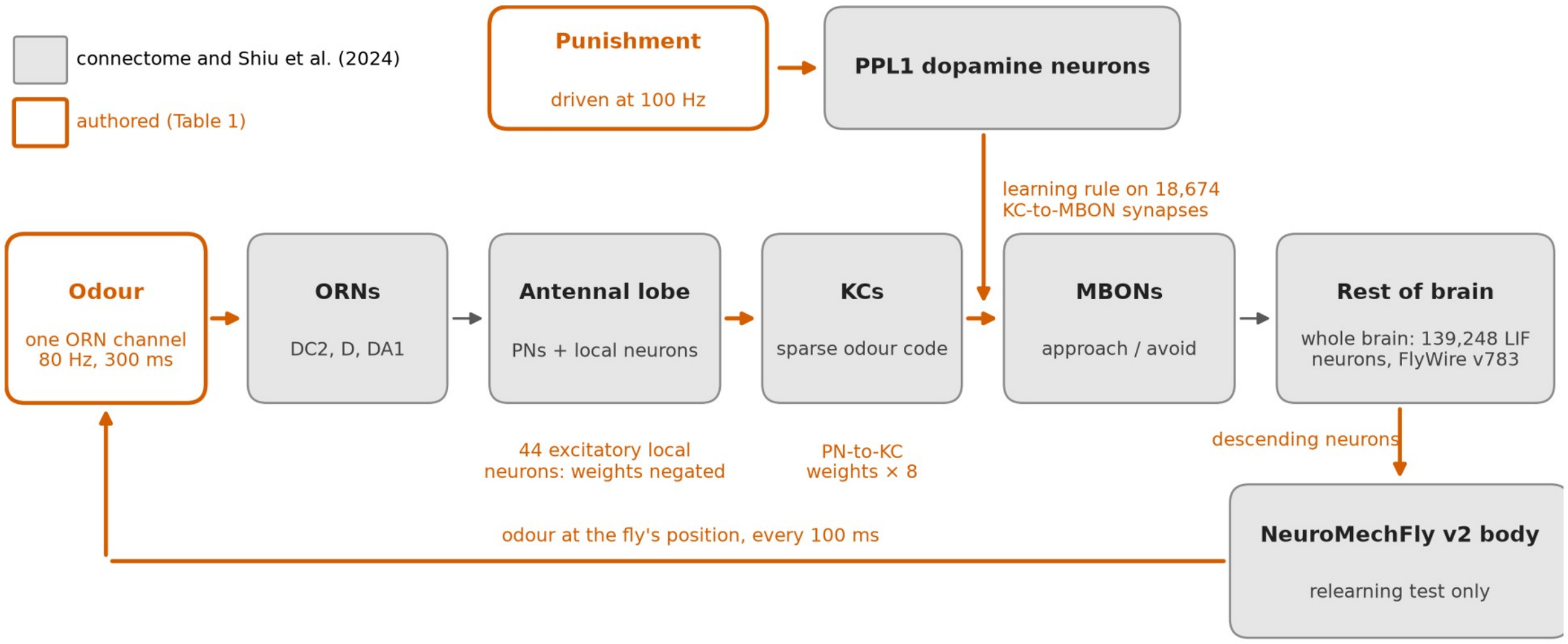

**Figure 1. The model: a whole-brain connectome with a small set of authored parts.**

Grey boxes come from the FlyWire connectome (v783) and the neuron model of Shiu et al. (2024); vermilion marks every part I authored (Table 1). An odour drives one class of olfactory receptor neurons (ORNs), whose signal passes through the antennal lobe to the Kenyon cells (KCs) and on to the mushroom body output neurons (MBONs). Punishment drives the PPL1 dopamine neurons, and their dopamine changes the synapses from KCs to MBONs under the learning rule. In the relearning test only, descending neurons drive the NeuroMechFly v2 body, and the odour at the fly's position is fed back to the ORNs every 100 ms. This panel is a schematic with no data.

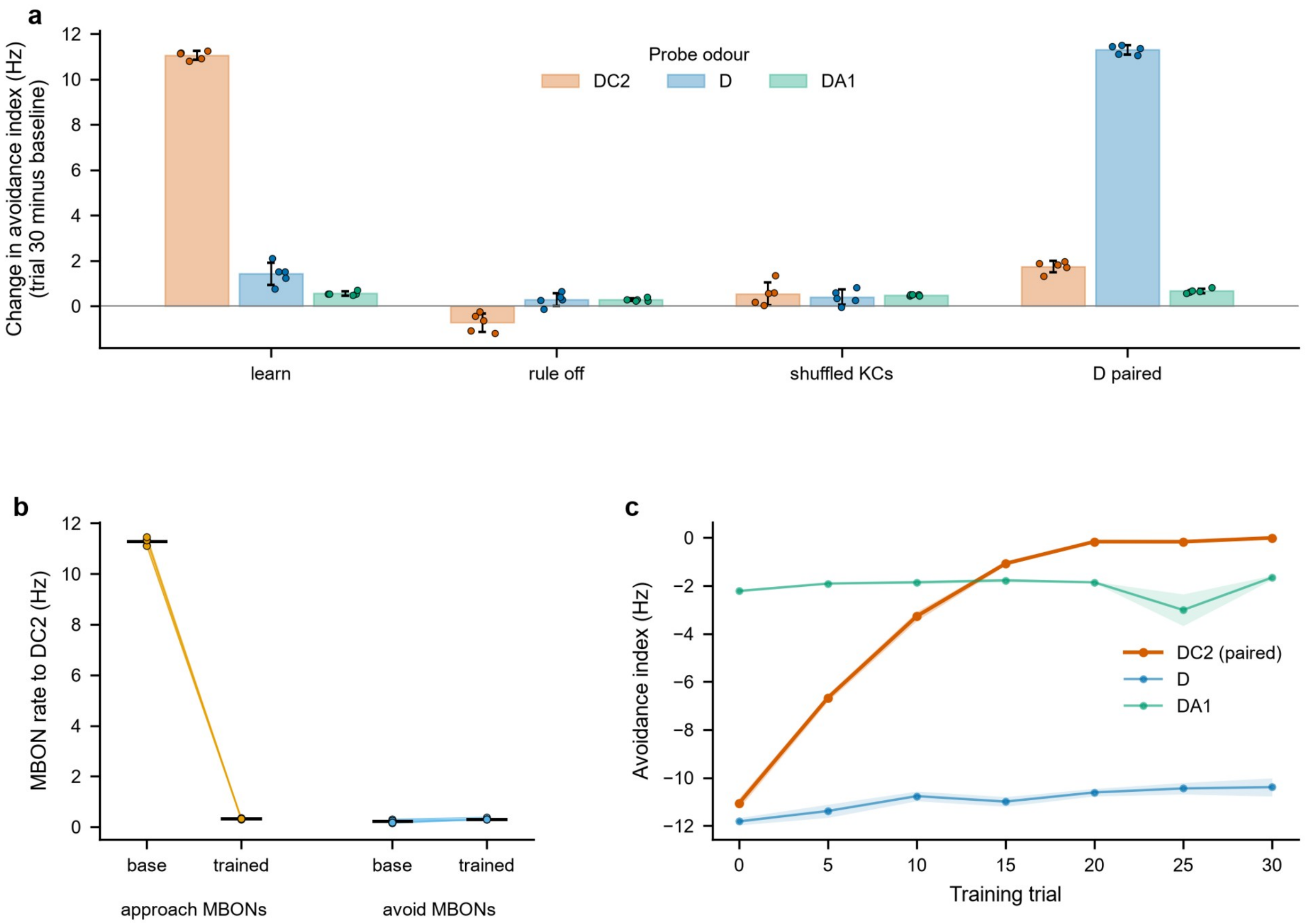

**Figure 2. Learning was specific to the paired odour.**

**(a)** Change in the avoidance index (mean avoid-MBON rate minus mean approach-MBON rate during a probe) from the baseline probe to the probe after trial 30, for each probe odour (DC2, D, DA1) in four arms: learn (DC2 paired with punishment), rule off (lesion), shuffled KCs (the rule reads each synapse's activity from a randomly reassigned KC), and D paired (the paired odour swapped). Bars = mean, points = seeds. In the learn arm, DC2 rose by 11.05 ± 0.20 Hz (5/5 seeds), D by 1.43 Hz and DA1 by 0.56 Hz; in the D-paired arm, D rose by 11.29 ± 0.21 Hz and DC2 by 1.74 Hz.

**(b)** Population rates of approach and avoid MBONs to DC2 in the learn arm, at baseline and after trial 30; lines join each seed, black ticks = mean. Approach MBONs fell from 11.3 to 0.3 Hz; avoid MBONs moved from 0.215 to 0.311 Hz.

**(c)** Avoidance index in the learn arm across training (probes every 5 trials), for the paired odour DC2 (bold) and for D and DA1 (light). Line = mean, band = SD.

n = 5 seeds (noise replicates of one simulated brain, not separate brain instances); error bars and bands show SD across seeds.

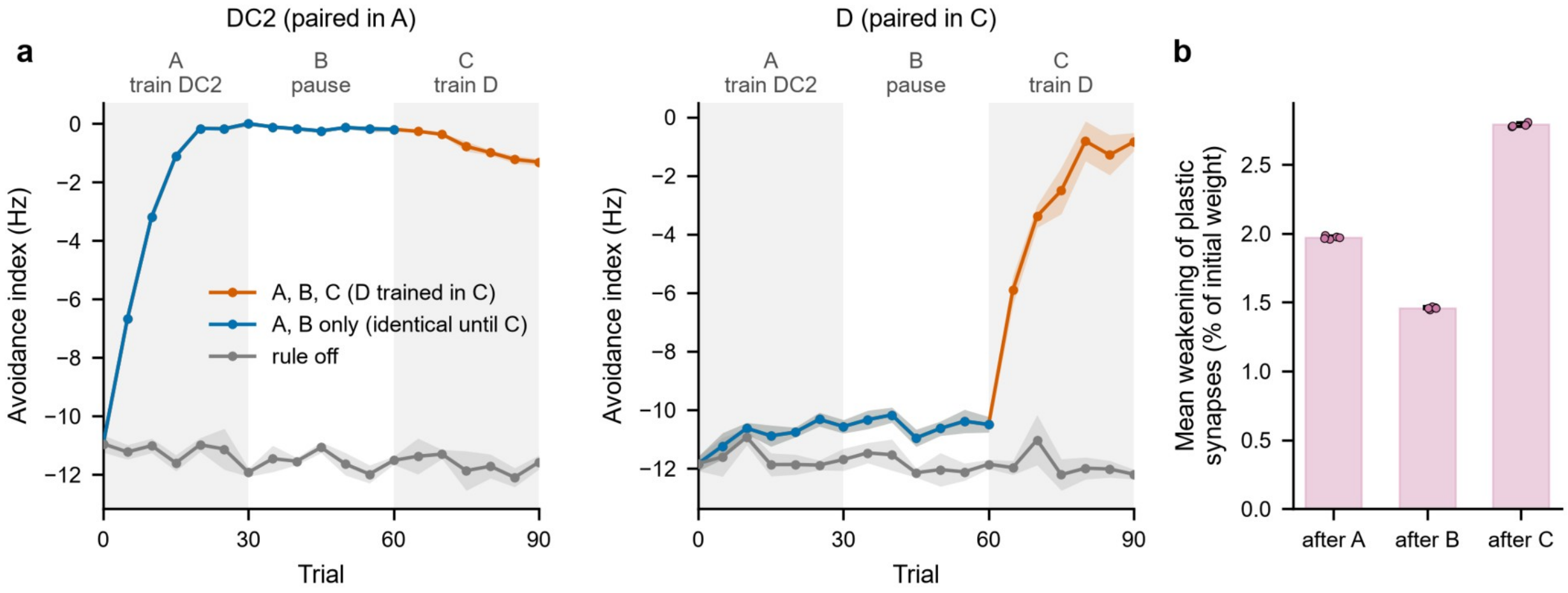

**Figure 3. A second memory formed on an already-trained brain.**

**(a)** Avoidance index to DC2 (left) and D (right) across three phases run on one brain without resetting its weights: A, DC2 paired with punishment (trials 1–30); B, a pause with no stimulus and no punishment (31–60); C, D paired with punishment (61–90). Orange, all three phases; blue, phases A and B only (identical to orange until C); grey, rule off in all phases. Line = mean, band = SD. Phase A raised DC2 by 10.96 ± 0.27 Hz; phase C raised D by 9.66 ± 0.45 Hz (85.5% of the 11.29 Hz reached on a naive brain), against −0.34 ± 0.17 Hz with the rule off. During C, DC2 lost 1.120 ± 0.147 Hz, against 0.195 ± 0.104 Hz over the equal stretch without training (phase B).

**(b)** Mean weakening of the plastic synapses from KCs to MBONs, as a percentage of their initial weight, at the end of each phase (arm with all three phases). Bars = mean, points = seeds. The weakening fell from 1.97% to 1.46% over phase B, as set by the recovery constant λ ($0.99^{30}$ = 74%), and rose to 2.79% after phase C.

n = 5 seeds (noise replicates of one simulated brain, not separate brain instances); error bars and bands show SD across seeds.

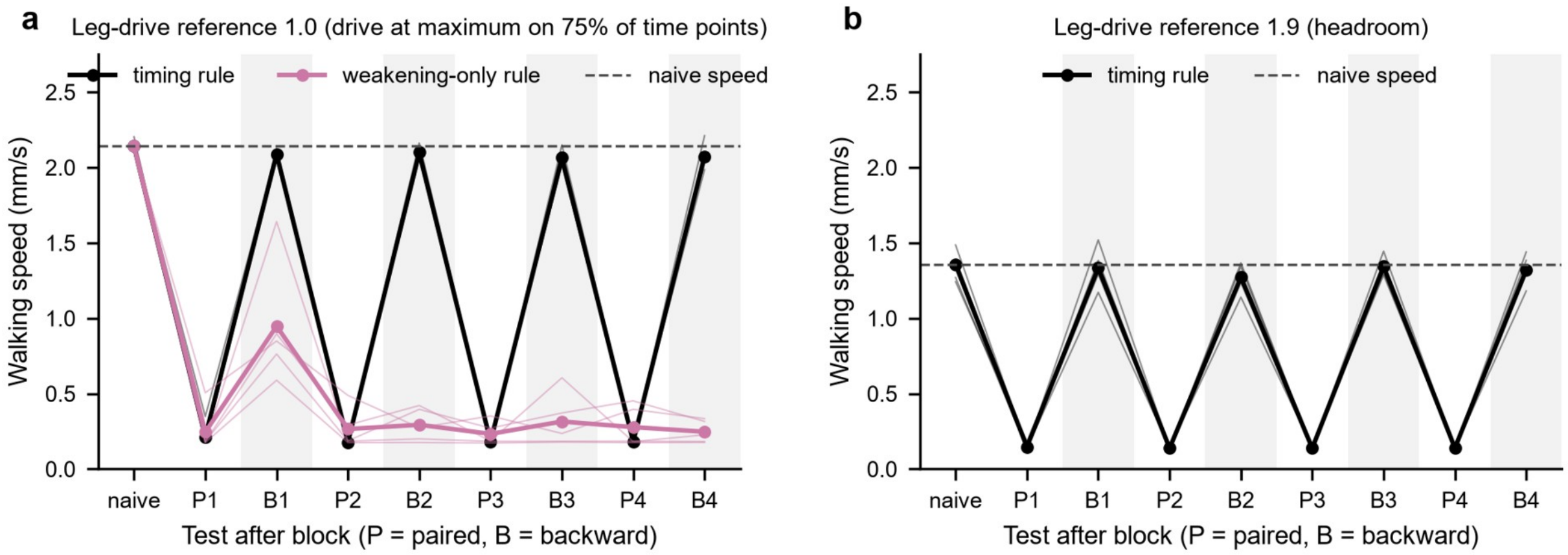

**Figure 4. Only the rule with timing relearned repeatedly.**

**(a)** Walking speed of the simulated fly in 60-tick test blocks: before training (naive) and after each of four cycles of a 200-tick paired block (P1–P4, dopamine after the odour) and a 200-tick backward block (B1–B4, shaded, dopamine before the odour). Black, timing rule; pink, rule that only weakens synapses. Thin lines = seeds, thick line = mean, dashed = naive speed (2.14 mm/s). At this setting of the descending-neuron reference rate (which sets the gain from firing rate to leg central-pattern-generator, CPG, amplitude; reference = 1.0), the drive was at its maximum on 75% of time points. The timing rule returned to 2.07–2.10 mm/s after every backward block (mean 97% of naive); the weakening-only rule recovered to 0.95 mm/s after B1 and never exceeded 0.32 mm/s afterwards.

**(b)** As in (a), timing rule only, with the descending-neuron reference rate raised to 1.9, which lowers the drive-to-amplitude gain and gives the readout more headroom before saturating; naive speed 1.355 mm/s (lower than in (a), since the same firing rate now drives less leg amplitude). The weakening-only rule was not run at this setting. Speed changed in the predicted direction at 8 of 8 transitions.

n = 5 seeds (noise replicates of one simulated brain, not separate brain instances); error bars and bands show SD across seeds.

**Table S1. Results that depend on the synapse floor: each claim at the canonical floor of 5 synapses and at a floor of 1 (the released model's connection set, the transmitter signs used elsewhere in this paper — Shiu et al.'s predicted signs, except the 44 antennal-lobe local neurons, whose outgoing sign is flipped — all authored constants unchanged).**

n = number of seeds (noise replicates of one brain) behind each value. Floor-5 and floor-1 columns use the same seeds (seed 0-2 means, or seed 0 alone), so they compare like with like; the last column gives the 5-seed floor-5 value quoted in the preprint text where it differs. Labels come from the preregistered labeller of the floor-1 recheck (HOLDS = same conclusion, SHIFTED = same sign, different size, FLIPS = conclusion changes).

| Claim (section of the text) | Floor 5 (canonical) | Floor 1 | Floor-1 / floor-5 | n seeds | Label | Floor 5 as quoted in text (n) |
|---|---|---|---|---|---|---|
| Paired DC2 shift, Hz (Results, specificity) | +10.96 | +29.25 | 2.67x | 3 (seeds 0-2) | SHIFTED | +11.05 ± 0.20 (5) |
| Unpaired D shift, Hz (Results, specificity) | +1.26 | +2.49 | 1.98x | 3 (seeds 0-2) | HOLDS | +1.43 (5) |
| Unpaired / paired ratio (Results, specificity) | 0.115 | 0.085 | - | 3 (seeds 0-2) | HOLDS | 12.9% (5) |
| Never-paired DA1 shift, Hz (Results, specificity) | +0.59 | +2.10 | 3.55x | 3 (seeds 0-2) | SHIFTED | +0.56 (5) |
| Lesion DC2 shift, Hz (Results, specificity) | −0.98 | −0.85 | 0.87x | 3 (seeds 0-2) | HOLDS | −0.73 (5) |
| Reassigned-KC control / learning, DC2 (Results, specificity) | 0.024 | 0.019 | - | 3 (seeds 0-2) | HOLDS | 4.9% (5) |
| Reversed arm: D shift, Hz (Results, specificity) | +11.17 | +23.77 | 2.13x | 3 (seeds 0-2) | SHIFTED | +11.29 ± 0.21 (5) |
| Reversed arm: DC2 shift, Hz (Results, specificity) | +1.61 | +1.58 | 0.98x | 3 (seeds 0-2) | HOLDS | +1.74 (5) |
| Approach MBONs to DC2, baseline → trained, Hz (Results, specificity) | 11.20 → 0.31 | 26.01 → 0.92 | 2.32x (baseline) | 3 (seeds 0-2) | SHIFTED | 11.3 → 0.3 (5) |
| Avoid MBONs to DC2, change, Hz (Results, specificity) | +0.07 | **+4.16** [4.11, 4.19, 4.19] | - | 3 (seeds 0-2) | **FLIPS** | 0.215 → 0.311, i.e. +0.10 (5) |
| Accumulation test, phase A: DC2 shift, Hz (Results, accumulation) | +11.11 | +29.05 | 2.61x | 1 (seed 0) | SHIFTED | +10.96 ± 0.27 (5) |
| Readout (avoidance index) retention over the pause (Results, accumulation) | 96.6% | 81.2% | - | 1 (seed 0) | SHIFTED | 98.2% (5) |
| Weight retention over the pause (Results, accumulation; set by λ) | 73.97% | 73.97% | - | 1 (seed 0) | HOLDS (pinned) | 74.0% (5) |
| Phase C D / naive D (Results, accumulation) | 82.1% | 91.0% | - | 1 (seed 0) | HOLDS | 85.5% (5) |
| DC2 change during D training / during pause (Results, accumulation) | 2.32 | **0.67** | - | 1 (seed 0) | **FLIPS** | 5.7 (5; ratio of means) |
| DC2 change kept at end of phase C (Results, accumulation) | 88.8% | 68.6% | - | 1 (seed 0) | SHIFTED (this value); FLIPS (interference claim overall) | 88% (5) |
| Lesion DC2, start → end of phase C, Hz (Results, accumulation) | −0.33 | −1.33 | - | 1 (seed 0) | HOLDS | −0.63 (5) |
| KCs active for DC2 / D / DA1; DC2-D Jaccard (Methods, odour) | .060 / .055 / .089; .057 | .061 / .057 / .113; .061 | - | 1 gate run | HOLDS | same (gate) |
| Shiu et al. (2024) benchmarks (Methods, neuron model) | 8 PASS, 3 PARTIAL, 2 FAIL of 13 | 4 PASS, 1 FAIL of 5 run (8 not run) | - | 5 | not labelled (different comparison, see note) | 8 / 3 / 2 of 13 (5) |
| Motor neuron MN9 at 50 Hz sugar, Hz (Methods, neuron model: the floor weakens the response to weak sugar input) | 2.0 | 15.2 (paper 19.4) | - | 5 | not labelled | - |

Note on the last two rows: eight of the thirteen benchmarks were not run against the model as Shiu et al. (2024) released it (their connection matrix at a floor of 1, their original transmitter signs); because this configuration differs from the canonical model in both floor and transmitter sign at once, it does not separate the two effects on the five benchmarks it does cover. Two of the eight not rerun are the ones that failed at the canonical floor of 5: the silencing of single neurons during sugar input at 50 Hz (Shiu et al., 2024, Fig. 1f-h), which failed because the control condition's response to sugar was itself nearly silent at this floor, and the comparison of how two classes of Johnston's organ neurons drive the aBN1 neuron (Shiu et al., 2024, Fig. 5g), whose failure remains unexplained.

**Table S2. Specificity and accumulation results at a projection-neuron-to-KC gain of 8 (canonical) and 6, all other settings unchanged.**

n = number of seeds (noise replicates of one brain). Shift = avoidance index at the last probe minus the baseline probe (Hz). Gain 8 is shown twice for the specificity test: as the 5-seed mean quoted in the text and as seed 0 alone, because the gain-6 specificity run has only seed 0. "Reassigned-KC control" is the shuffle arm of the specificity test (Methods), in which the rule reads each synapse's activity from a randomly reassigned KC.

## Specificity test

| Quantity | Gain 8, 5 seeds (n = 5) | Gain 8, seed 0 (n = 1) | Gain 6, seed 0 (n = 1) |
|---|---|---|---|
| Paired DC2 shift, Hz | +11.05 ± 0.20 | +10.90 | +5.10 |
| Unpaired D shift, Hz | +1.43 ± 0.49 | +1.51 | -0.01 |
| D / DC2 shift | **12.9%** (ratio of the 5-seed means; per-seed range 7.0-18.9%) | **13.9%** | **-0.2%** |
| Never-paired DA1 shift, Hz | +0.56 ± 0.09 | +0.53 | +0.18 |
| DA1 / DC2 shift | 5.0% | 4.9% | 3.6% |
| Lesion arm DC2 shift, Hz | -0.73 ± 0.41 | -1.10 | -0.62 |
| Lesion / learning | -6.6% | -10.1% | -12.2% |
| Reassigned-KC control DC2 shift, Hz | +0.54 ± 0.51 | +0.58 | -0.14 |
| Reassigned-KC control / learning | 4.9% | 5.3% | -2.7% |
| Approach MBONs to DC2, baseline -> trained, Hz | 11.28 -> 0.32 | 11.13 -> 0.30 | 5.54 -> 0.44 |
| Share of baseline approach drive removed | 97.1% | 97.3% | 92.1% |
| Avoid MBONs to DC2, baseline -> trained, Hz | 0.215 -> 0.311 | 0.296 -> 0.370 | 0.296 -> 0.296 |
| Reversed arm (D paired) | D +11.29 ± 0.21 | D +11.04 | not run |
| Gate: KCs active for DC2 / D / DA1 (one gate run per gain) | .060 / .055 / .089 | same gate | .055 / .050 / .061 |
| Gate: DC2-D KC Jaccard; silent channels; spontaneous KCs | 0.057; 0; 0 | same gate | 0.060; 0; 0 |

## Accumulation test

| Quantity | Gain 8 (n = 5) | Gain 6 (n = 3) |
|---|---|---|
| Phase A DC2 shift, Hz | +10.96 ± 0.27 | +5.08 ± 0.14 |
| Phase C D shift, Hz | +9.66 ± 0.45 | +2.58 ± 0.15 |
| Phase C D / naive D | 85.5% (naive = reversed arm of the specificity test, gain 8, 5 seeds) | not computable (no reversed arm at gain 6) |
| Share of remaining approach drive to D removed in phase C | 90.0% | 72.7% |
| DC2 change during the pause, Hz | -0.20 ± 0.10 | -0.07 ± 0.05 |
| DC2 change during D training, Hz | -1.12 ± 0.15 | -0.56 ± 0.04 |
| Ratio, D training / pause (ratio of means) | 5.7 | 7.7 |
| DC2 change kept at end of phase C | 88.0% ± 0.7 | 87.6% ± 0.3 |
| Approach MBONs at end, DC2 / D, Hz (baseline) | 1.56 / 1.10 (11.20 / 12.41) | 0.89 / 0.97 (5.53 / 4.32) |
| Mean weakening of plastic synapses after A / pause / C | 1.97% / 1.46% / 2.79% | 1.56% / 1.15% / 2.24% |
| Weight retention over the pause (set by $\lambda = 0.01$) | 74.0% (SD 0) | 74.0% (SD 0) |
| Readout retention over the pause | 98.2% ± 0.9 | 98.6% ± 1.0 |
| Lesion arm, D shift in phase C, Hz | -0.34 ± 0.17 | -0.63 ± 0.55 |
| Lesion arm, DC2 start -> end, Hz | -0.63 ± 0.44 | -0.47 ± 0.19 |

Values are mean ± SD across seeds (sample SD) where n > 1.